\documentclass[journal,12pt,onecolumn,draftclsnofoot]{IEEEtran}
\usepackage[top=0.75in,bottom=0.90in,right=0.75in,left=0.75in]{geometry}
\makeatletter
\def\ps@headings{%
\def\@oddhead{\mbox{}\scriptsize\rightmark \hfil \thepage}%
\def\@ adversarynhead{\scriptsize\thepage \hfil \leftmark\mbox{}}%
\def\@oddfoot{}%
\def\@ adversarynfoot{}}
\makeatother
\ifCLASSINFOpdf
\else

\fi

\usepackage{epsfig}
\usepackage{array}
\newcolumntype{L}[1]{>{\raggedright\let\newline\\\arraybackslash\hspace{0pt}}m{#1}}
\newcolumntype{C}[1]{>{\centering\let\newline\\\arraybackslash\hspace{0pt}}m{#1}}
\newcolumntype{R}[1]{>{\raggedleft\let\newline\\\arraybackslash\hspace{0pt}}m{#1}}
\usepackage{amsmath}
\usepackage[short]{optidef}
\usepackage{amsthm} 
\usepackage{mathrsfs}
\newcommand{\bc}{\begin{center}}
\newcommand{\ec}{\end{center}}
\newcommand{\be}{\begin{equation}}
\newcommand{\ee}{\end{equation}}
\usepackage{flexisym}
\usepackage{algorithmicx}
\usepackage{algorithm}
\usepackage{soul}
\usepackage{algpseudocode}

\usepackage{amsmath}

\newcommand{\bnu}{\begin{enumerate}}
\newcommand{\enu}{\end{enumerate}}
\usepackage{cite}
\usepackage{url}
\usepackage{breqn}
\usepackage{lipsum}
\usepackage{mathtools}
\usepackage{amsmath}
\usepackage{caption}
\usepackage{subcaption}
\usepackage{soul}
\usepackage{blindtext, graphicx}
\usepackage{cuted}
\usepackage{color}

\usepackage{amsfonts}
\usepackage{lipsum}
\usepackage{cuted}
\newtheoremstyle{case}{}{}{}{}{}{:}{ }{}

\begin{document}
\newgeometry{top=1in,bottom=0.90in,right=0.75in,left=0.75in}
\title{Enhanced IoT Batteryless D2D Communications Using Reconfigurable Intelligent Surfaces}
\author{Shakil Ahmed,~\IEEEmembership{Student Member,~IEEE}, Mohamed Y. Selim,~\IEEEmembership{Senior Member,~IEEE}, and Ahmed E. Kamal,~\IEEEmembership{Fellow,~IEEE}\vspace*{-0.75cm}
\thanks{ Shakil Ahmed, Mohamed Y. Selim, and Ahmed E. Kamal are with the Department of Electrical and Computer Engineering, Iowa State University, Ames, Iowa, USA. (email: \{shakil, myoussef, kamal\}@iastate.edu).
}}


\maketitle


\begin{abstract}
Recent research on reconfigurable intelligent surfaces (RIS) suggests that the RIS panel, containing passive elements, enhances channel performance for the internet of things (IoT) systems by reflecting transmitted signals to the receiving nodes.
This paper investigates RIS panel assisted-wireless network to instigate minimal base station (BS) transmit power in the form of energy harvesting for batteryless IoT sensors to maximize bits transmission in the significant multi-path environment, such as urban areas.
Batteryless IoT sensors harvest energy through the RIS panel from external sources, such as from nearby BS radio frequency (RF) signal in the first optimal time frame, for a given time frame.
The bits transmission among IoT sensors, followed by a device-to-device (D2D) communications protocol, is maximized using harvested energy in the final optimal time frame.
The bits transmission is at least equal to the number of bits sampled by the IoT sensor.
We formulate a non-convex mixed-integer non-linear problem to maximize the number of communicating bits subject to energy harvesting from BS RF signals, RIS panel energy consumption, and required time.
We propose a robust solution by presenting an iterative algorithm.
We perform extensive simulation results based on the 3GPP Urban Micro channel model to validate our model.
\end{abstract}

\section{Introduction}
The reconfigurable intelligent surface (RIS), which may be mounted on buildings or tall objects, has emerged as a 6G candidate technology enabling tunable scattering of incident signals in the channel propagation model \cite{1}.
Next-generation wireless communications require energy-efficient operation while supporting high-capacity communications.
RIS panels may help support these two requirements by controlling the reflection of the transmitted signal to the receiving entities, hence, achieving robust channels \cite{0}.
The RIS panel is a metasurface that consists of many low-cost passive elements, which control the reflection angle and attenuation of the incident signals according to the internal state.

Device-to-device (D2D) communications allow IoT sensors to communicate directly without going through a base station (BS) and may be used to support the excessive number of the next generation wireless nodes, which is one of the application domains of the 5G wireless networks, i.e., massive machine type communication.
Challenges arise when the internet of things (IoT) sensors are batteryless and required to harvest energy for operation.
However, harvesting energy from external sources, such as solar or wind energy, is not always a viable solution.
Therefore, adopting a scenario where IoT sensors harvest energy from nearby BS radio frequency (RF) signals to support D2D communications is an alternative solution that can be made practical of the RF signals and is enhanced by using RIS panels.
Deploying the RIS panels can significantly enhance IoT sensors' direct communication even for complex channel conditions by exploiting the RIS's passive elements reflection property.
RIS panels consume energy due to reflecting the incident signals to the receiver.

Our objective in this paper is to investigate how the RIS panels can facilitate energy harvesting from BS RF signals and D2D communications.
Many RIS-related works were published \cite{2,RISSurvey1,RISSurvey2} providing only a general overview of RIS panels in many areas such as channel estimation and modeling, optimization, resource allocation, localization, etc.
In addition, these surveys address the references by comparing the RIS panel with other technologies, such as a relay.
For example, the authors in \cite{2} designed the RIS panel-assisted energy-efficient system maximization problem.
The authors investigated a downlink scenario to optimize the RIS panel phase shifters, subject to minimum quality of service requirements.
Many papers address the usage of RIS panels to enhance D2D communications \cite{RISD2D1, RISD2D2, RISD2D3, RISD2D4}.
Meanwhile, other papers considered using RIS panels to increase the amount of harvested energy either for the RIS panel itself or the end sensors \cite{RISEH1, RISEH2, RISEH3}.

The authors in \cite{RISD2D1} studied RIS panel usage for enhancing the D2D communications underlying system performance.
They studied joint power control and passive beamforming optimization.
The authors in \cite{RISD2D2} focused on the RIS panel-assisted single-cell uplink communications with D2D communications sharing the same spectrum to minimize the interference.
In \cite{RISD2D3}, the authors investigated the RIS panel-assisted D2D communications under $Nakagami$-$m$ fading.
For energy harvesting, the authors in \cite{RISEH1} used the RIS panel to enhance wireless power transfer from the BS to non-line-of-sight (NLOS) links to maximize the minimum harvested power.
The authors in \cite{RISEH2} and \cite{RISEH3} proposed energy harvesting to empower the RIS panel.
In our previous efforts in \cite{SA}, we showed a trade-off between RIS panels with active and passive elements energy consumption and energy harvested from BS RF signals.

\begin{figure}[!h]
\centering
\includegraphics[width=5.0in]{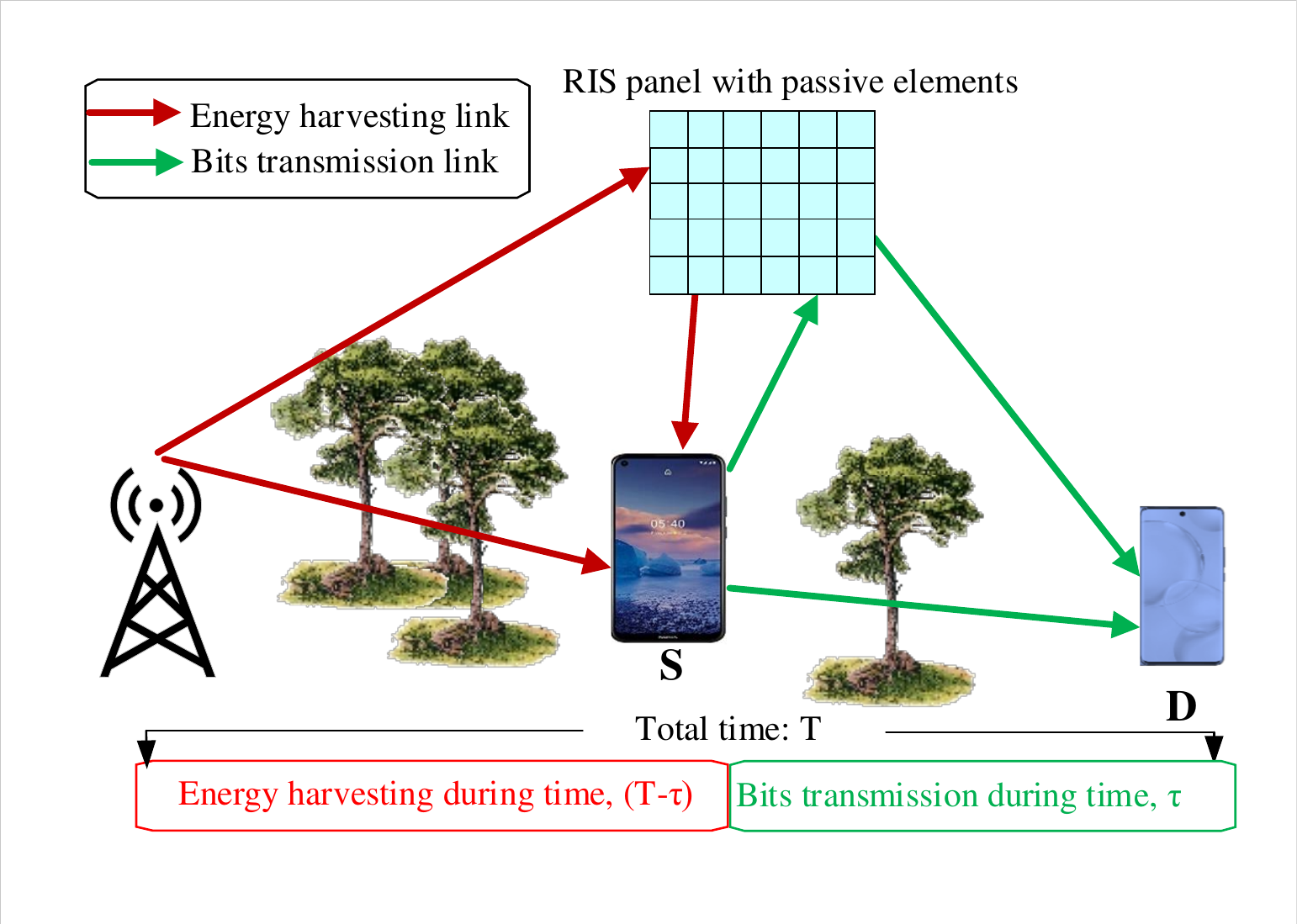}
\caption{System model }
\label{System_Model1}
\end{figure}

The contribution of this paper is as follows:
We propose a multi-purpose use of the RIS panel, powered by energy harvested from BS RF signals, for a fixed duration time frame divided into two optimal time slots, as shown in Fig.~\ref{System_Model1}.
It includes 1) \textit{energy harvesting from BS RF signals:} IoT sensor, $S$ harvests energy during the first time slot. We introduce an energy efficiency factor determining the amount of energy harvested from BS RF signals,
2) \textit{enhanced D2D communications:} Using this harvested energy, $S$ enhances D2D communications to another IoT sensor, $D$ in the remaining time slot.
We formulate an optimization problem to maximize the number of bits transmitted to $D$, including constraints, such as energy harvesting threshold, RIS panel passive elements energy consumption, and the required time at both phases.
The formulated non-convex mixed-integer non-linear problem (MINLP) is challenging to solve.
We, therefore, propose an iterative algorithmic solution.
We adopt various methods, such as Dinkelbach algorithm and Taylor series approximation, to develop the solution algorithm.
We present the results of the proposed model validated by the 3GPP channel model \cite{3GPP}.


\section{System and Channel Models} \label{SystemModel}
\subsection{System Parameters}
We consider a wireless network, where BS and IoT sensors are equipped with a single antenna.
We adopt all possible links without/with the assistance of the RIS panel between the communicating nodes.
The passive elements of the RIS panel reflect the incident BS RF signals from transmitting nodes towards destinations by modeling the propagation environment into a desirable form.
Our proposed model has two operational phases.
RIS panel, having $N$ discrete passive elements, is used in both phases.
In phase 1, a batteryless IoT sensor, $S$, harvests energy from BS RF signals during $(T-\tau)$ time, where $T$ is the total time frame.
In phase 2, $S$ communicates with $D$ using D2D communications and the energy harvested during $\tau$ time and assisted by the RIS panel.
$S$ samples data at a rate $S_r$, which is needed to transmit at least the amount of sampled data, $S_r(T-\tau)$ during $\tau$.
We summarize the used parameters in Table~I.

\subsection{Phase 1: Energy Harvesting During $(T-\tau)$}
\subsubsection{Passive elements properties of RIS panel}
The passive elements of RIS panel have a size smaller than wavelength, i.e., $\frac{\lambda}{5} \times \frac{\lambda}{5}$ \cite{3} and $\frac{\lambda}{8} \times \frac{\lambda}{8}$ \cite{4}, where $\lambda$ is the wavelength.
Due to this size, the RIS panel scatters the incident signals in a controlled manner to pre-determined directions.
We express the RIS panel properties to $S$ as a diagonal matrix as $\boldsymbol{\Theta } = \mathrm {diag} \left ({\alpha_1 e^{j \theta _{1}}, \alpha_2 e^{j \theta_2}, .., \alpha_N e^{j \theta _{N}} } \right)$, where $\alpha_i \in [0,1] $ is the fixed amplitude reflection coefficient and $\theta_i$ is the phase-shift of elements, respectively.
The passive elements require energy for operation and signal reflection, enabling energy harvesting at $S$ in phase 1.
Let $m$ be the number of contributing elements from $N$ passive elements during BS RF signals phase energy harvesting.
Let $y$ be the required power by each contributing element to stay functional.
During $(T-\tau)$, the energy required to power the contributing passive elements is $m y (T-\tau)$, which is also harvested from BS RF signals.

\subsubsection{Channel gain} \label{CG_I}
We encounter different channel gains between BS-to-$S$, BS-to-RIS and RIS-to-S, denoted as $h_{bs}$, $\mathbf {h}_{\mathrm {br}} \in \mathbb{C}^{N \times 1} $ and $\mathbf {h}_{\mathrm {rs}} \in \mathbb{C}^{N\times 1}$, respectively.
The received signals from BS to $S$ is:
\be
\begin{aligned}
s= ({h}_{bs}+ \mathbf {h}_{\mathrm {rs}}^{ { \mathrm {\dagger}}} \boldsymbol{\Theta } \mathbf {h}_{\mathrm {br}}) \sqrt {P_s} s + n
\end{aligned}
\ee
where $s$ is the unit-power information signals and $P_s$ is the BS transmit power.
$\mathbf {h}_{\mathrm {rs}}^{ { \mathrm {\dagger}}}$ is a conjugate transpose matrix.
These gains are modeled as Rayleigh fading with $n \sim \mathcal{CN}(0,1)$.
We consider the frequency flat quasi-static block fading model.
Thus, the destination knows them correctly.
In this model, the channel is assumed to be coherent during the transmission phases while it changes from cycle to cycle.
For any given $\boldsymbol{\Theta}$, we have the gains as follows:
\be
\begin{aligned}
\mathbf {h}_{\mathrm {rs}}^{ { \mathrm {\dagger}}} \boldsymbol{\Theta} \mathbf {h}_{\mathrm {br}}= \sum_{n=1}^N \alpha_n e^{j \theta_n} [\mathbf {h}_{\mathrm {rs}}^{\dagger}]_n [\mathbf {h}_{\mathrm {br}}]_n
\end{aligned}
\ee

\textcolor{black}{Using $\mathbf {h}_{\mathrm {rs}}^{ { \mathrm {\dagger}}} \boldsymbol{\Theta} \mathbf {h}_{\mathrm {br}}$, the maximum rate is achieved.
Phase shift $\theta_n$ is selected in the sum to be the same phase of $h_{bs}$.
$\theta_n$ is, therefore, expressed \cite{1} as $\textcolor{black}{\theta_n=\arg(h_{bs}) - \arg([\mathbf {h}_{\mathrm {rs}}^{\dagger}]_n [\mathbf {h}_{\mathrm {br}}]_n)}.$ }
\begin{table}[]
\caption{List of symbols used in the paper}
\begin{tabular}{|c|l|c|l|}
\hline
Symbol & Description & Symbol & Description \\ \hline
$\sigma^2$ & Noise & $W$ & Bandwidth \\ \hline
$\alpha_i$ & Reflection coefficient & $B^{\tau}$ & Transmitted bits \\ \hline
$T$ & Total time & $\theta, \phi$ & RIS phase shift \\ \hline
$P_s$ & BS power & $S_r$ & Sampling rate \\ \hline
$\lambda$ & Wavelength & $N$ & RIS elements \\ \hline
$h_{sd}$ & $S$-$D$ channel gain & $m$ & RIS elements (Phase 1) \\ \hline
$h_{bs}$ & BS-$S$ channel gain & $k$ & RIS elements (Phase 2) \\ \hline
$\mathbf{h_{rs}}$ & RIS-$S$ channel gain & $\zeta$ & Efficiency factor \\ \hline
$\mathbf{h_{rd}}$ & RIS-$D$ channel gain & $d_1\!-\!d_4$ & Constants \\ \hline
$\mathbf{h_{br}}$ & BS-RIS channel gain & $e_m$ & Min. harvested energy \\ \hline
$\tau$ & D2D time & $y$ & Power for each element \\ \hline
$\mathbf{\Theta, \Phi}$ & Diagonal matrix & $(T-\tau)$ & Energy harvesting time \\ \hline
$\mathbf{h_{sr}}$ & $S$-RIS channel gain & $\Im_1, \Im_2$ & Slack variables \\ \hline
$[.]^*$ & Feasible point & $\lambda_1, \lambda_2$ & Constants \\ \hline
$s$ & Received signal & $\mathbf{e}$ & Harvested energy \\ \hline
$\mathbf{h_1}, \mathbf{h_2}$ & Channel gains & $W$ & Bandwidth \\ \hline
\end{tabular}
\end{table}

\subsubsection{Energy harvesting from BS RF signals}
The power-in-power-out wireless charging technology expresses energy harvesting from BS RF signals.
The required energy harvesting at $S$ to operate is expressed as follows:
\be
\begin{aligned}
\label{RF_EH}
\mathbf{e} = \zeta (T-\tau)P_s \left({ | {h}_{bs}|+ m | [\mathbf {h}_{\mathrm {rs}}^{\dagger}]_{n} \, [\mathbf {h}_{\mathrm {br}}]_{n} | } \right)^{2}
\end{aligned}
\ee
where $\zeta \in [0,1)$ is the energy efficiency factor.
We focus on D2D communications between the IoT sensors, $S$, and $D$ with energy harvesting from BS RF signals. 
\vspace*{-0.4 cm}
\subsection{Phase 2: D2D Communications During $\tau$}
\subsubsection{Passive elements properties of RIS panel}
The passive elements require energy for operation and signal reflection, enabling D2D communications in phase 2.
The RIS panel properties are also represented by the diagonal matrix as $\boldsymbol{\Phi } = \mathrm {diag} \left ({\alpha _1 e^{j \phi _{1}}, \alpha _2 e^{j \phi_2}, \ldots, \alpha _N e^{j \phi_{N}} }\right)$, where $\phi_i$ is the phase-shift of elements.
Let $k$ be the number of contributing elements from $N$ passive elements.
$k$ requires $k y \tau$ energy to stay functional during $\tau$ time, where $y$ is the required power by each passive contributing element.

\subsubsection{Channel gains}
We adopt a similar channel condition from Section~\ref{CG_I}.
Thus, the channel gains between $S$-to-$D$, $S$-to-RIS panel and RIS panel-to-$D$ are $h_{sd}$, $\mathbf {h}_{\mathrm {sr}} \in \mathbb{C}^{N \times 1} $ and $\mathbf {h}_{\mathrm {rd}} \in \mathbb{C}^{N\times 1}$, respectively.
For any given $\boldsymbol{\Phi}$, we have gains as $\textcolor{black}{\mathbf {h}_{\mathrm {rd}}^{ { \mathrm {\dagger}}} \boldsymbol{\Theta} \mathbf {h}_{\mathrm {sr}}= \sum_{n=1}^N \alpha_n e^{j \theta_n} [\mathbf {h}_{\mathrm {rd}}^{\dagger}]_n [\mathbf {h}_{\mathrm {sr}}]_n}$.
Phase shift, $\phi_n$ is selected in the sum to be the same phase of $h_{sd}$.
$\phi_n$, therefore, is expressed \cite{1} as $\phi_n=\arg(h_{sd}) - \arg([\mathbf {h}_{\mathrm {rd}}^{\dagger}]_n [\mathbf {h}_{\mathrm {sr}}]_n).$

\subsubsection{Transmissions of IoT measurement bits}
The maximum number of bits transmitted from $S$ to $D$ during $\tau$ time slot, without/with RIS-assisted D2D communications, is obtained using (\ref{RF_EH}) as follows:
\be
\begin{aligned}
\label{Rate_D2D}
B^{\tau}=&\max _{\phi_{1},\ldots,\phi_{N}} W \tau \log _{2} \left ({1 + \textcolor{black}{\frac {\frac{e}{\tau} | {h}_{sd}+ \mathbf {h}_{\mathrm {dr}}^{ { \mathrm {\dagger}}} \boldsymbol{\Phi } \mathbf {h}_{\mathrm {rs}}|^{2}}{\sigma ^{2}} }}\right) \\
=& W \tau \log _{2} \left ({1 + \frac {\frac{e}{\tau}\left({| {h}_{sd}|+ k |\mathbf {h_1}|_n }\right)^{2}}{\sigma ^{2}} }\right)
\end{aligned}
\ee
where $W$ is bandwidth of the carrier frequency and $|\mathbf {h_1}| = | [\mathbf {h}_{\mathrm {rd}}^{\dagger}]_{n} \, [\mathbf {h}_{\mathrm {rs}}]_{n} |$.
$S$ has the following operation to transmit the number of bits:
$S$ samples at a rate of $S_r$. This results in at least the $S_r(T-\tau)$ bits, e.g., $S_r(T-\tau) \leq B^{\tau}$.
\vspace*{-0.4 cm}
\section{Problem Formulation}
Recall that $T$ is the total time, $\tau$ is the time for D2D communications, and $(T-\tau)$ is the time during energy harvesting from BS RF signals.
We find optimal $\tau$ for a given $T$ to allocate the optimal time slot at each phase.
We formulate an optimization problem to maximize the number of transmitting bits, which is at least being equal to the number of sampled bits as a function of $\tau, m,$ and $k$ as follows:
\begin{subequations}\label{ob.2}
\begin{align}\label{ob.2.1}\
&{\mathop {\max }\limits_{\tau, m, k}
}\ {\textcolor{black}{W \tau \log _{2} \left ({1 + \frac {\frac{\mathbf {e} }{\tau}\left({| {h}_{sd}|+ k | \mathbf {h_1}| }\right)^{2}}{\sigma ^{2}} }\right)}}\\
&\text{s.t.}\ e_m \leq \zeta (T - \tau) P_s \left({ | {h}_{bs}|+ m | \mathbf {h_2}| } \right)^{2} \label{ob.2.c2.}\\
& m y (T-\tau) \leq (N-m) \mathbf {e} (T-\tau) \label{ob.2.c3}\\
& k y \tau \leq (N-k) \mathbf {e} \tau \label{ob.2.c6}\\
& S_r (T-\tau) \leq W \tau \log_2 \left( 1+ \frac{\frac{\mathbf {e} }{\tau} \left({| {h}_{sd}|+ k | \mathbf {h_1}| }\right)^{2}}{\sigma^2} \right) \label{ob.2.c4}\\
& \tau < T \label{ob.2.c5}
\end{align}
\end{subequations}

where $\mathbf {h_2}= | \mathbf {h}_{\mathrm {rs}}^{\dagger}]_{n} \, [\mathbf {h}_{\mathrm {br}}]_{n} $.
Note (\ref{ob.2}) is non-convex due to coupling variables in (\ref{ob.2.1})~-~(\ref{ob.2.c4}).
We define the objective function in (\ref{ob.2.1}) maximizing bits transmission during the D2D phase for a given time frame, where harvested energy from BS RF signals is used from (\ref{RF_EH}).
We set a threshold of energy harvesting in (\ref{ob.2.c2.}) as $e_m$.
$S$ harvests energy from BS RF signals above a threshold that allows $S$ to transmit bits to $D$ during $\tau$.
Energy consumed by RIS passive elements required during RF signals harvesting is described in (\ref{ob.2.c3}).
On the other hand, energy consumed by RIS elements required during the D2D communications phase is described in (\ref{ob.2.c6}).
Equation (\ref{ob.2.c4}) guarantees that $S$ harvests enough energy that allows it to transmit all sampled bits at a rate of $S_r$.
Time constraint during energy harvesting and D2D communications is described in (\ref{ob.2.c5}).
To deal with the nature of the non-convex MINLP in (\ref{ob.2}), Section~\ref{PS} proposes a heuristic solution approach to this problem.

\section{Proposed Solution} \label{PS}
\vspace{- 0.17 cm}
We see that (\ref{ob.2}) is a non-convex problem.
It is challenging to tackle energy harvesting in (\ref{ob.2.c2.}), bit transfer above sampling rate in (\ref{ob.2.c4}), and energy consumption constraints in (\ref{ob.2.c3})~-~(\ref{ob.2.c6}), due to the coupling of the optimizing variables.
To efficiently solve the optimization problem in (\ref{ob.2}), we transform it into more tractable equivalent \textcolor{black}{three} sub-problems.
Firstly, we optimize the RIS panel passive elements used during energy harvesting, given passive elements used in D2D communications and their corresponding time.
This step introduces slack variables, applies some feasible points, and uses a first-order Taylor series approximation to make the optimization problem convex.
Secondly, we retain the optimal RIS passive elements used during energy harvesting to optimize the RIS passive elements used during D2D communications.
We add the slack variables and Dinkelbach algorithm due to the fractional nature of the constraint.
We also apply the first-order Taylor series to make the optimization problem convex at some feasible points.
Finally, using the optimal elements obtained from the above steps, we optimize the time required for D2D communications.
Again, we apply the first-order Taylor series to make the optimization problem convex at some feasible points.
An iterative approach follows those processes.
In the following parts, the three sub-problems are investigated in turn.

\subsection{Sub Problem 1}
Given D2D communications time, $\tau$, and contributing passive elements, $k$, we optimize contributing passive elements of RIS panel during energy harvesting from BS RF signals phase, $m$, therefore, making the problem more tractable.
The newly formulated optimization problem is written as follows:
\vspace{-0.35 cm}
\begin{subequations}\label{sub_1}
\begin{align}\label{sub_1..}\
&{\mathop {\max }\limits_{m}
}\ {\textcolor{black}{W \tau \log _{2} \left ({1 + d_1 \mathbf {e} }\right)}}\\
&\text{s.t.}\ e_m \leq (T-\tau)P_s \left({| {h}_{bs}|+ m |\mathbf {h_2}| }\right)^{2} \label{sub_1.c2}\\
& m y \leq (N-m) \mathbf {e} \label{ob.2.c3.}\\
& S_r (T-\tau) \leq W \tau \log_2 \left( 1+ d_1 \mathbf {e} \right) \label{ob.2.c4..}
\end{align}
\end{subequations}
where $d_1=\frac{\left({| {h}_{\mathrm {sd}}|+ k | \mathbf {h_1}| }\right)^{2}}{\tau \sigma^2}$.
\textcolor{black}{We introduce a slack variable $\Im_1$ to tackle the objective function in (\ref{sub_1..}).} Thus, we have the reformulated optimization problem as follows:
\begin{subequations}\label{sub_1.2f.}
\begin{align}\label{sub_1.2.}\
&{\mathop {\max }\limits_{m,\Im_1}
}\ {\textcolor{black}{W} \tau \log_2 (1+ d_1 \Im_1)}\\
&\text{s.t.}\ \mathbf {e} \leq \Im_1 \label{sub_1.2.c1} \\
& e_m \leq \zeta (T-\tau)P_s \left({| {h}_{s}|+ m |\mathbf {h_2}| }\right)^{2} \label{sub._1.c2}\\
& m y \leq (N-m) \mathbf {e} \label{ob.2.c3.}\\
& S_r (T-\tau) \leq W \tau \log_2(1+ d_1 \Im_1) \label{ob.2.c10}
\end{align}
\end{subequations}

Note that (\ref{ob.2.c3.}) is not convex yet due to coupling variables.
Using a feasible point and (\ref{RF_EH}), the reformulated (\ref{ob.2.c3.}) is:
\be
\begin{aligned}
\label{ob.2.c3...}
1 \leq \frac{(|h_{bs}|+ m |\mathbf {h_2})| \bigg( c_1 h^{*}-\frac{c_2mm^*}{m^*+c_3}h^{*} \bigg ) }{m}
\end{aligned}
\ee
where $c_1=\frac{\zeta^2 (T-\tau) }{y}$ and $h^{*}=(|h_{bs}|+ m^* |\mathbf {h_2}|)$.
$c_3$ is constant.
However, this is still not a convex problem due to its fractional nature.
We apply the Dinkelbach algorithm and approximate it to a convex problem.
We briefly explain the Dinkelbach algorithm in the following (interested readers can find more information in \cite{Dinkelbach67}).
Here, $f(r)\!=\frac{P(r)}{Q(r)}$ is described as $f(r)\!\!=\!\!P(r)\!-\!\lambda_1 Q(r)$ under all convex constraints, where $\lambda_d$ is a constant.
This value is iteratively updated by $\lambda_{1,j+1}\!=\!\frac{p_{r_j}}{Q_{r_j}}$, where $j$ is the iterative index.
This process guarantees convergence, and hence, the locally optimal solution is achieved.
We reformulate (\ref{ob.2.c3...}) as follows:
\be
\begin{aligned}
\label{eq_3}
1 \! \leq \! (|h_{bs}|+ m |\mathbf {h_2}|) \bigg( \! c_1 h^{*} \!- \!\frac{c_2mm^*}{m^*+c_3}h^{*}\! \bigg ) - \lambda_1^i m
\end{aligned}
\ee
where $\lambda_1^i$ is a new numerical value that is iteratively updated.
Similarly, we linearize (\ref{sub._1.c2}) as follows:
\be
\begin{aligned}
\label{eq_4}
e_m \! \leq \! \zeta (T\!-\!\tau)P_s \! \left({| {h}_{bs}|+ m |\mathbf {h_2}| }\right) \left({| {h}_{\mathrm {bs}}|+ \alpha m^* |\mathbf {h_2}| }\right)
\end{aligned}
\ee

Thus, the optimization problem is written as follows:
\begin{align} \label{sub_1.2f}
& \underset{m,\Im_1}{\text{max}} \textcolor{black}{W \tau} \log_2 (1+ d_1 \Im_1), \\
& \text{s.t.}\ (\ref{sub_1.2.c1}), (\ref{ob.2.c10}), (\ref{eq_3}), (\ref{eq_4}) \nonumber
\end{align}

Now (\ref{sub_1.2f}) is efficiently solved via existing optimization techniques.
\textcolor{black}{Since the optimal solution to (\ref{sub_1.2f}) is not necessarily binary in general, we apply the first order Taylor series expansion at feasible point $\Im_1^*$ in the objective function of (\ref{sub_1.2.}) and (\ref{ob.2.c10}).}
\begin{subequations} \label{sub_1_5}
\begin{align}
&{\mathop {\max }\limits_{m, \Im_1}
}\ { \textcolor{black}{W \tau } \bigg ( \frac{d_1 (\Im_1-\Im_1^*)}{(1+ d_1 \Im_1^*)}+\log _{2}(1+ d_1 \Im_1^*) \bigg )}\\
&\text{s.t.}\ S_r (T-\tau) \! \leq \! \bigg ( \frac{d_1 (\Im_1-\Im_1^*)}{(1+ d_1 \Im_1^*)}+\log _{2}(1+ d_1 \Im_1^*) \bigg ) \\
& (\ref{sub_1.2.c1}), (\ref{eq_3}), (\ref{eq_4}) \nonumber
\end{align}
\end{subequations}

Now (\ref{sub_1_5}) is a convex problem.

\subsection{Sub Problem 2}
Given D2D communications time, $\tau$ and using the optimal passive elements, $m$ during energy harvesting, we optimize the RIS panel elements, $k$ during D2D communications phase. Therefore, the optimization problem is given by:
\begin{subequations}
\begin{align}\label{sub_2}\
&{\mathop {\max }\limits_{k}
}\ {\textcolor{black}{W \tau \log _{2} \left ({1 + \frac {\frac{\mathbf {e} }{\tau}\left({| {h}_{sd}|+ k |\mathbf {h_1}|}\right)^{2}}{\sigma ^{2}} }\right)}}\\
&\text{s.t.}\ k y \leq (N-k) \mathbf {e} \label{156}\\
& S_r (T-\tau) \leq W \tau \log_2 \left( 1+ \frac{\frac{\mathbf {e} }{\tau} \left({| {h}_{sd}|+ k | \mathbf {h_1}| }\right)^{2}}{\sigma^2} \right) \label{ob.2.c4.}
\end{align}
\end{subequations}
\begin{figure*}
\centering
\begin{subfigure}[b]{0.32\textwidth}
\centering
\includegraphics[width=\textwidth]{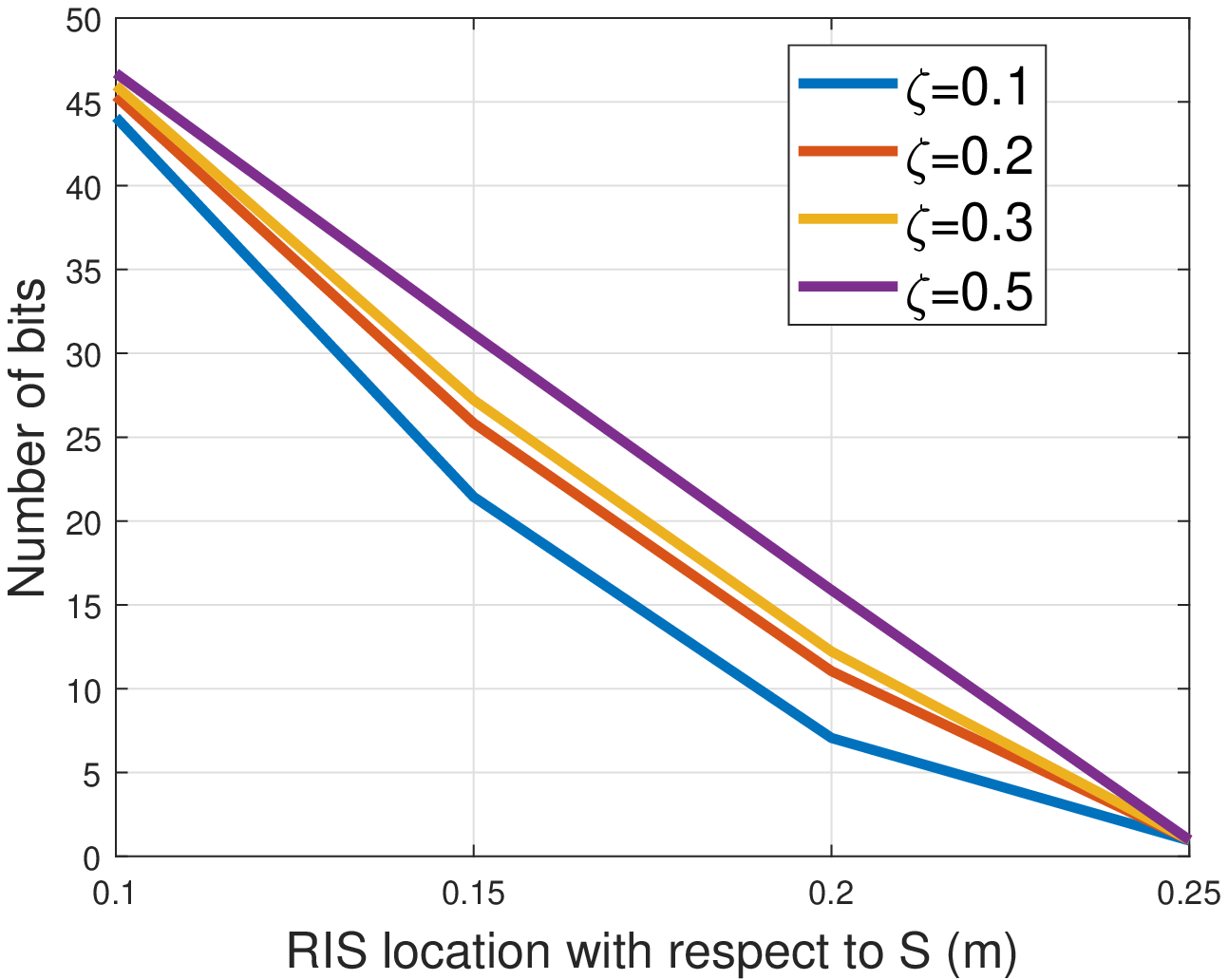}
\caption{Bits changing with RIS locations}
\label{P_1}
\end{subfigure}
\hfill
\begin{subfigure}[b]{0.32\textwidth}
\centering
\includegraphics[width=\textwidth]{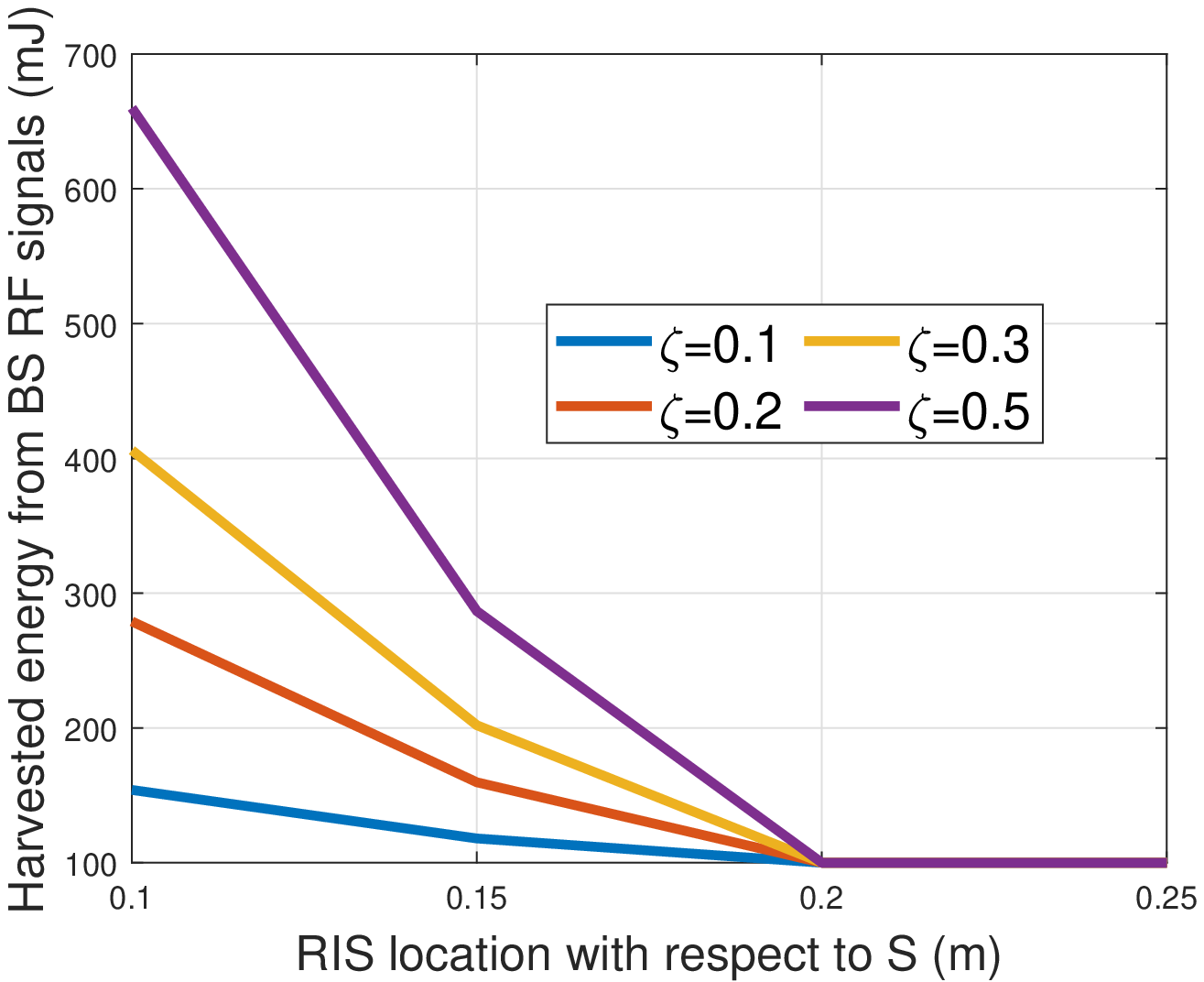}
\caption{Energy harvesting vs. RIS locations}
\label{P_2}
\end{subfigure}
\hfill
\begin{subfigure}[b]{0.32\textwidth}
\centering
\includegraphics[width=\textwidth]{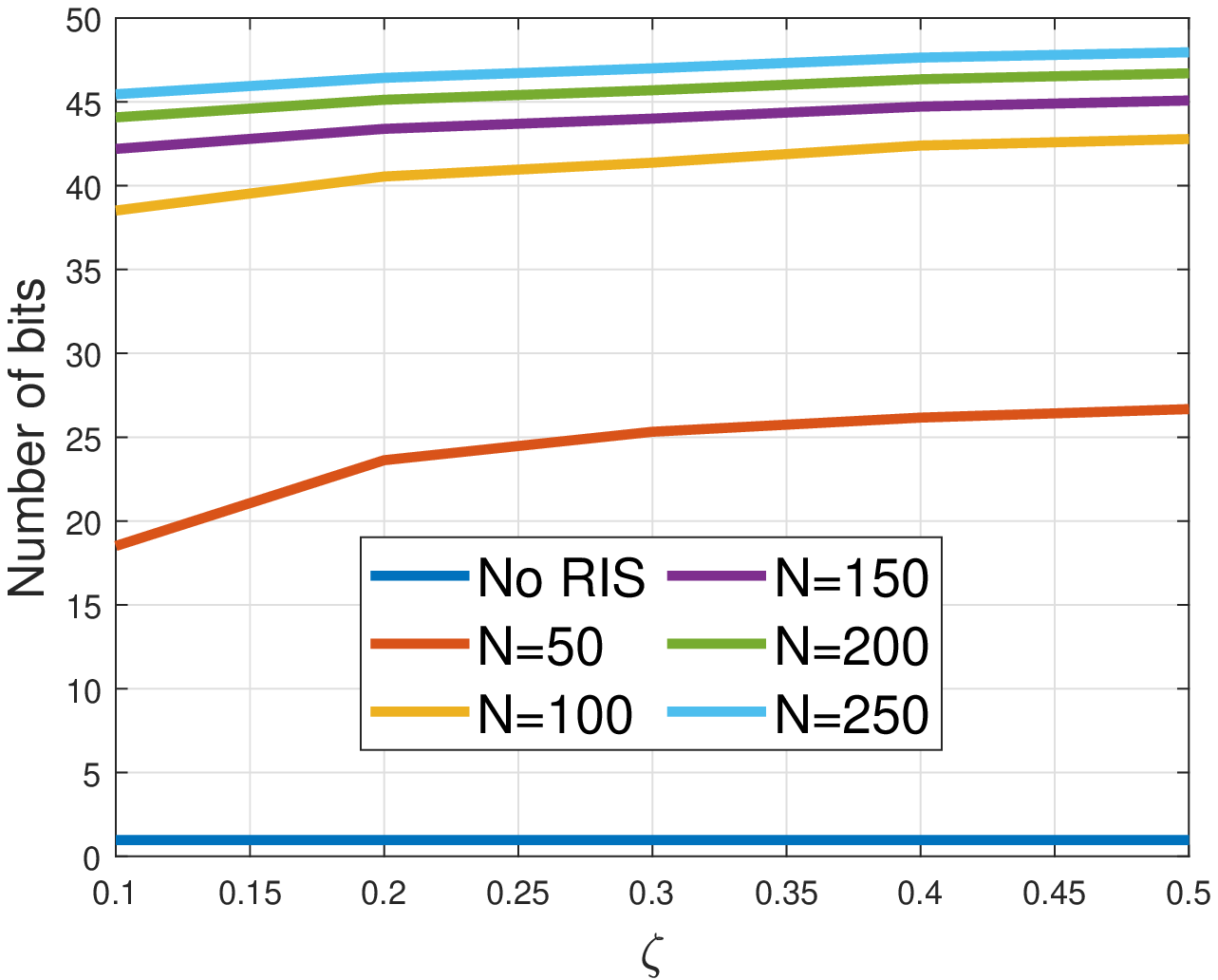}
\caption{Bits vs. $\zeta$}
\label{P_3}
\end{subfigure}
\hfill
\begin{subfigure}[b]{0.33\textwidth}
\centering
\includegraphics[width=\textwidth]{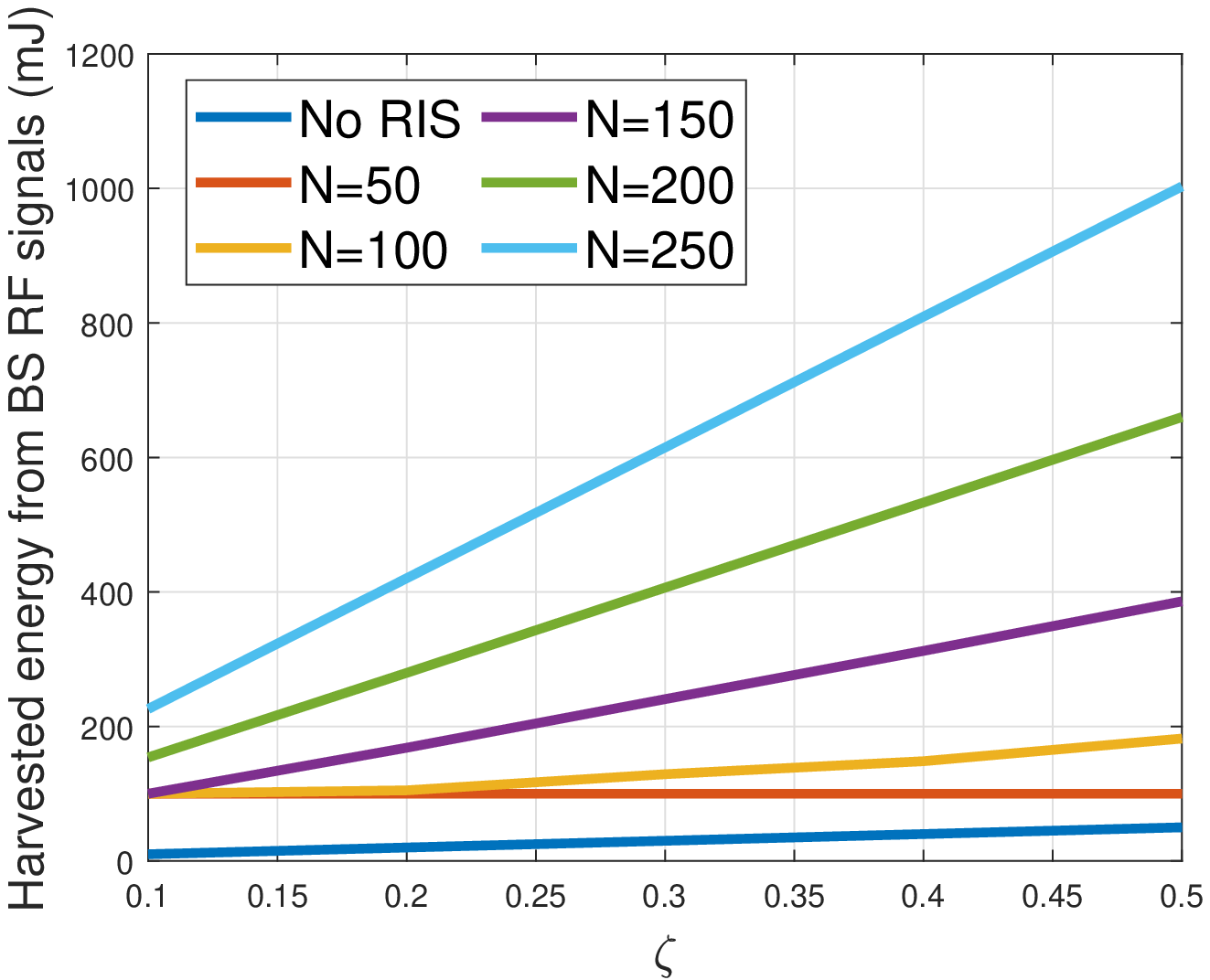}
\caption{Harvested energy with increasing $\zeta$}
\label{P_4}
\end{subfigure}
\hfill
\begin{subfigure}[b]{0.32\textwidth}
\centering
\includegraphics[width=\textwidth]{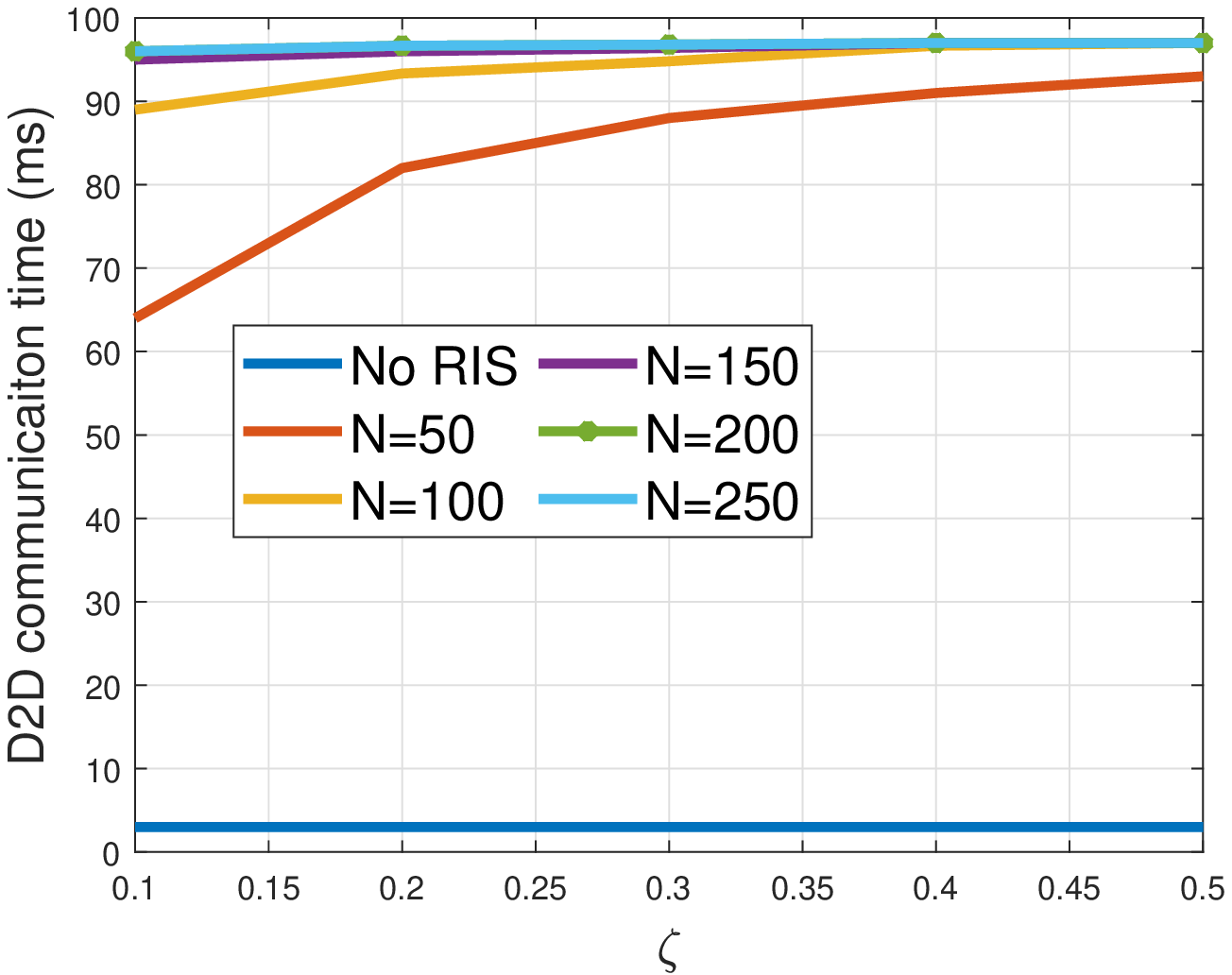}
\caption{D2D communications time vs $\zeta$.}
\label{3P}
\end{subfigure}
\hfill
\begin{subfigure}[b]{0.33\textwidth}
\centering
\includegraphics[width=\textwidth]{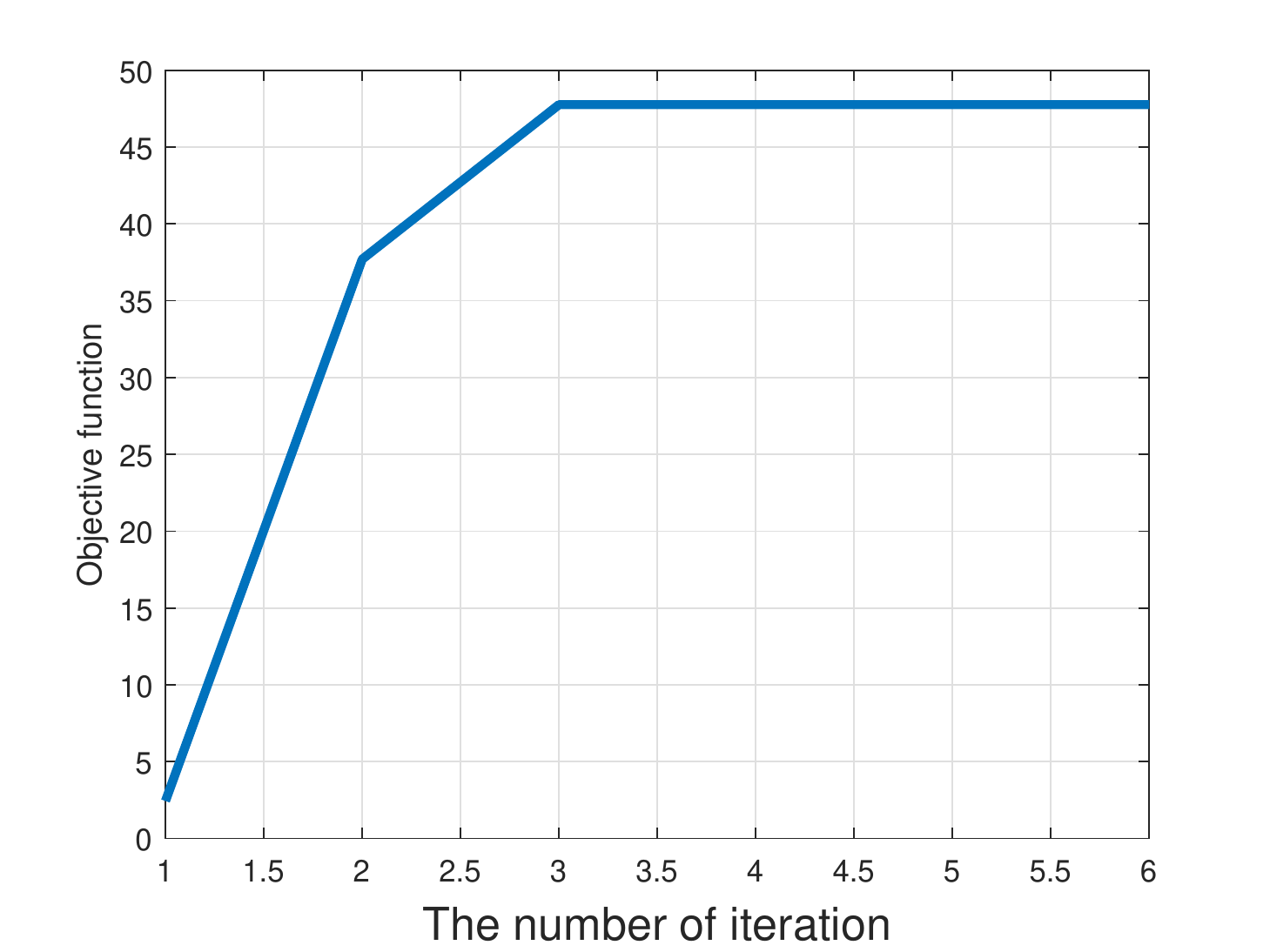}
\caption{Convergence performance}
\label{Fig_Conv}
\end{subfigure}
\caption{The effect of RIS panel (a)~-~(b) by changing the RIS locations, (c)~-~(d) by varying energy efficiency factor, $\zeta$, (e) Optimal time slot allocation during D2D communications, and (f) convergence of the algorithm}
\label{1P}
\end{figure*}

We tackle the objective function as follows: We introduce a variable $\Im_2$ in the nominator of $\log(.)$ function.
The optimization is reformulated as follows:
\begin{subequations}
\begin{align}\label{sub_2.1}\
&{\mathop {\max }\limits_{k, \Im_2}
}\ {\textcolor{black}{W \tau }\log _{2} \left ({1 + \frac {\mathbf {e} }{\tau \sigma ^{2}} \Im_2}\right)}\\
&\text{s.t.}\ \Im_2 \leq \left({| {h}_{\mathrm {sd}}|+ k |\mathbf {h_1} | }\right)^{2} \label{12} \\
& S_r (T-\tau) \leq W \tau \log_2 \big( 1+ \frac{\mathbf {e} }{\tau \sigma^2} \Im_2 \big)\label{eq_18c} \\
& (\ref{156}) \nonumber
\end{align}
\end{subequations}

We see that (\ref{156}) is not a convex problem yet.
We adopt Dinkelbach algorithm to approximate as convex.
\be
\begin{aligned}
\label{eq_12}
1 \leq \bigg ( \frac{N \mathbf {e} \zeta}{y} \bigg (1- \lambda_2^i k \bigg ) -\frac{\mathbf {\mathbf {e} } \zeta}{ y} \bigg)
\end{aligned}
\ee

Now we apply the first order Taylor series approximation at some feasible point $\Im_2^*$ in the objective function of (\ref{sub_2.1}) and also in (\ref{eq_18c}).
\begin{subequations} \label{sub_213}
\begin{align}\label{sub_2.2}\
&{\mathop {\max }\limits_{k, \Im_2}
}\ { \textcolor{black}{W \tau } \bigg ( \frac{\mathbf {e} (\Im_2-\Im_2^*)}{\tau \sigma^2 (1+ \mathbf {e} \Im_2^*)}+\log _{2}(1+\frac{\mathbf {e} \Im_2^*}{\tau \sigma^2}) \! \! \bigg )}\\
&\text{s.t.}\ \Im_2 \leq \left({| {h}_{sd}|+ k |\mathbf {h_1}|}\right) \left({| {h}_{sd}|+ k^* |\mathbf {h_1}|}\right) \\
& S_r \! \leq \! \frac{W \tau \log_2 \bigg ( \!\! \frac{\mathbf {e} (\Im_2 - \Im_2^*)}{\tau \sigma^2(1+\mathbf {e} \Im_2^*)} \! + \! \log_2 \! \bigg( \! 1 \! + \! \frac{\mathbf {e} \Im_2^*}{ \tau \sigma^2} \bigg ) \!\!\! \bigg )}{(T-\tau)}\\
&(\ref{eq_12}) \nonumber
\end{align}
\end{subequations}

Finally, (\ref{sub_213}) is a convex problem.

\subsection{Sub Problem 3}
Using the optimal number of passive elements during energy harvesting, $m$, and D2D communications, $k$, the optimal D2D communications time, $\tau$ for a given $T$ optimization problem is reformulated as follows:
\begin{subequations}\label{sub_3}
\begin{align}\label{sub_3.1}\
&{\mathop {\max }\limits_{\tau}
}\ {\textcolor{black}{W} \tau \log _{2} \left ({1 + \frac{d_3(T-\tau)}{\tau}}\right) }\\
&\text{s.t.}\ e_m \leq d_4 (T-\tau) \label{sub_3.1.c2}\\
& S_r (T-\tau) \leq W \tau \log_2 \left( 1+ \frac{d_3(T-\tau)}{\tau} \right) \label{sub_3_c1} \\
& \tau < T \label{ob.2.c5.}
\end{align}
\end{subequations}
where $d_3=\zeta P_s\left({| {h}_{bs}|+ |\mathbf {h_1}| }\right)^{2}$ and $d_4=\frac{d_3 \left({| {h}_{sd}|+ k |\mathbf {h_2}| }\right)^{2} }{\sigma^2}$.
Both (\ref{sub_3.1.c2}) and (\ref{ob.2.c5.}) are linear and convex. However, (\ref{sub_3}) is not a convex problem due to (\ref{sub_3.1}) and (\ref{sub_3_c1}).
Similarly, we reform (\ref{sub_3_c1}) as follows:
\be
\begin{aligned}
\label{eq:sub_3.2}
1 \! \leq \! W \tau \! \bigg [ \!\! \log_2(d_3(T-\tau)+\tau) \! - \! \log_2(\tau) \bigg] \!\! + \! S_r (\tau - T)
\end{aligned}
\ee

Finally, we rewrite (\ref{sub_3.1}) by applying the first-order Taylor series approximation and reformulating the optimization problem.
Hence, (\ref{sub_3}) is reformulated as follows:
\begin{align} \label{eq_last}
& \underset{\tau}{\text{max}} W \tau \bigg [ \log_2(d_3(T-\tau)+\tau) \! - \! \log_2(\tau) \bigg] \\
& \text{s.t.}\ (\ref{sub_3.1.c2}), (\ref{ob.2.c5.}), (\ref{eq:sub_3.2}) \nonumber \nonumber
\end{align}

Note that (\ref{eq_last}) is a convex problem.
At every iteration, the feasible points, including $m$, $k$, and $\tau$, are updated, and this process
continues until it reaches convergence \cite{Dinkelbach67, Ref_Conv}.
We summarize the proposed solution in
an algorithm as follows:
\begin{algorithm}
\caption{}
\label{alg:algorithm_sum}
\begin{algorithmic}[1]
\State $\bold{Input:}$ Parameters, $T$, $N$, $\sigma^2$, $\zeta$, $P_s$, $S_r$, and $e_m$
\State $\bold {Optimization}$:
\Repeat
\State Find optimal element used in energy harvesting from BS RF signals for given $k$ and $\tau$ using (\ref{sub_1_5})
\State Find optimal elements used in D2D communications $k$ using (\ref{sub_213}) for given $\tau$ and optimal $m$
\State Find optimal time during D2D communications using (\ref{eq_last}) for optimal $m$ and $k$ from previous state
\Until{convergence}
\end{algorithmic}
\end{algorithm}

\section{Simulation Results}
In this section, we \textcolor{black}{present our results based on the optimization problem and its solution}.
The BS, $S$, $D$, and RIS panel are deployed at fixed locations.
We consider a quasi-static channel while calculating the channel gain based on Table B.1.2.1-1 from 3GPP channel model \cite{3GPP} with 3GHz carrier frequency.
To perform the simulation, we use GAMS, which solves MINLP \cite{gams}.
We consider $T=100$ ms, $N \in \lbrace 50, 100, 150, 200, 250 \rbrace$, $\sigma^2=-94$ dBm, $\zeta=[0,1)$, $P_s=1$ W, $e_m=0.1$ mJ, and $S_r=1 s^{-1}$.


In Fig.~\ref{P_1} and Fig.~\ref{P_2}, we show the network performance when we vary the RIS panel locations.
The bits transmission and energy harvesting reduce when the distance between RIS and $S$ increases.
This is because the location of $S$ plays the most critical role by being part of both energy harvesting from BS RF signals and the D2D communications.
Thus, a greater distance between the $S$ and RIS results in lower performance.
A higher energy efficiency factor provides a higher number of bits transmission because $S$ harvests more energy when $\zeta$ is higher.
In Fig.~\ref{P_1}, the number of transmitted bits is very close for various $\zeta$ when RIS is located 0.1 m far from $S$.
In Fig.~\ref{P_2}, we see a significant difference in harvested energy when RIS is located 0.1 m far for $S$.
This is because the transmitted signals from $S$ to $D$ may have different channel conditions.

Fig.~\ref{P_3} and Fig.~\ref{P_4}, we show the performance in terms of the number of bits transmission and harvested energy for various passive elements and compare when the system does not deploy the RIS panel.
A fewer RIS panel passive elements reduce the number of micro-controllers used for RIS; eventually, it becomes less expensive with less hardware complexity.
Fig~\ref{P_3} shows a higher number of passive elements enhances the number of bits transmission.
Fig.~\ref{P_4} shows the higher amount of energy is harvested from BS RF signals when we choose a higher number of passive elements and efficiency factors.
In both cases, the bits transmission and energy harvesting are at the lowest side when the model does not deploy any RIS panel, as shown in Fig.~\ref{P_3} and Fig.~\ref{P_4}, respectively.
An obstacle between $S$-to-$D$ and the BS-to-$S$ significantly weakens the received signals.
Fig.~\ref{3P}, D2D communications time does not change significantly when the model adopts a higher number of RIS elements, even for the higher power-efficiency factor.
Due to the greater number of passive elements, all possible links are reflected by the RIS panel to the IoT sensors. The time for the D2D communications phase between $S$ and $D$ increases with the distance between communicating nodes, i.e., decreasing channel gain.
On the other hand, the remaining time is utilized for energy harvesting from BS RF signals phase.
$S$ transmits the significant number of bits when the system has no RIS panel. D2D communications time is the least compared to the system deploys RIS panel.
Fig.~\ref{Fig_Conv} shows that the convergence of the optimization problem is fast and occurs after the third iteration.

\section{Conclusion}
We develop a framework to maximize the number of bits transmission for D2D communications without/through the RIS panel for batteryless IoT sensors.
We address the power constraint of batteryless IoT sensors and the RIS panel passive elements by harvesting energy from BS RF signals.
The BS RF signals powered-IoT sensor is responsible for D2D communications to another IoT sensor.
We consider that the transmitting nodes transmit the RF signals to the receiving nodes using all possible links.
We formulate an optimization problem to maximize the number of communicated bits subject to energy harvested from BS RF signals, RIS panel passive elements energy consumption, and their corresponding time frame.
The energy harvesting from BS RF signals and D2D communications occurs in different time slots while the duration of the total time frame is fixed.
The optimization problem is a non-convex MINLP.
We propose a robust solution to the problem by presenting an iterative algorithm. We also perform simulation results based on the 3GPP channel model to validate our model.

\end{document}